# Micromagnetic modeling of Terahertz oscillations in an antiferromagnetic material driven by spin-Hall effect

V. Puliafito[1], R. Khymyn[2], M. Carpentieri[3], B. Azzerboni[1], V. Tiberkevich[4], A. Slavin[4], G. Finocchio[5]

[1] Department of Engineering, University of Messina, 98166 Messina, Italy

[2] Department of Physics, Gothenburg University, 40530 Gothenburg, Sweden

[3] Department of Electrical and Information Engineering, Politecnico of Bari, 70125 Bari, Italy

[4] Department of Physics, Oakland University, 48309 Rochester, MI, United States

[5] Department of Mathematical and Computer Sciences, Physical Sciences and Earth Sciences, University of Messina, 98166 Messina, Italy

**Abstract**

The realization of THz sources is a fundamental aspect for a wide range of applications. Over different approaches, compact THz oscillators can be realized taking advantage of dynamics in antiferromagnetic (AFMs) thin films driven by spin-Hall effect. Here we perform a systematic study of these THz oscillators within a full micromagnetic solver based on the numerical solution of two coupled Landau-Lifshitz-Gilbert-Slonczewski equations, for the case of ultra-thin films, i.e. when the Néel temperature of an AFM is substantially reduced. We have found two different dynamical modes depending on the strength of the Dzyaloshinskii-Moriya interaction (DMI). At low DMI, a large amplitude precession is excited where both the magnetizations of the sublattices are in a uniform state and rotate in the same direction. At large enough DMI, the ground state of the AFM becomes non-uniform and the antiferromagnetic dynamics is characterized by ultrafast domain wall motion.

I. INTRODUCTION

Terahertz (THz) radiation covers the range of frequencies from 300GHz to 3THz, between microwaves and infra-red, corresponding to wavelengths ranging from 1000 to 100 μm [1,2]. Since a wide variety of lightweight molecules emits in this range of the electromagnetic spectrum, THz were intensely investigated by astronomers and chemists in the past [3,4]. However, THz oscillations have turned out to be very promising in many other fields, such as biomedicine [5], defense and security [6], material science [7], industrial non-destructive testing [8], and information and communication technology (ICT) [9, 10]. THz sources can be realized with quantum cascade lasers [11], solid state devices [12], however the development of compact nano-sized electrical generators and receivers of THz signals represents a key-challenge of the modern technology. With the experience maturated after decades of research on the generation and manipulation of GHz-frequency dynamics in ferromagnetic materials [13, 14, 15, 16, 17, 18], development of high quality AFM materials for several applications [19, 20, 21, 22, 23], and proof of concept of antiferromagnetic memories [24, 25, 26, 27] driven by the spin-Hall effect (SHE) [28], research is now combining this know-how focusing on the development of AFM-based oscillators for application in 4G and 5G telecommunication systems [29, 30, 31, 32, 33, 34, 35]. Up to now, there is no experimental proof of AFM-based oscillators (ASHO), and all the theoretical studies are considering two sub-lattices with their magnetizations antiferromagnetically coupled [36] and their dynamics is studied by solving two Landau-Lifshitz-Gilbert-Slonczewski (LLGS) equations [37] within the macrospin approximation [29, 31, 32].

Besides, in the above papers the Néel temperature of the AFM thin films was assumed to be much higher than the operational temperature of the oscillator. It is, however, known, that ultra-thin films of the magnetically ordered materials exhibit strong reduction of the phase transition temperature. Thus, the 1-5 nm thick film of nickel oxide (NiO) can become paramagnetic already at the room temperature [38]. Therefore, below we are interested in the case when the Néel temperature is close but still above the operational temperature. In this regime, the exchange stiffness constant is substantially reduced as compare to the values of a bulk crystal. In contrast, the influence of magnetic anisotropy for thin films can be stronger than in the bulk, because of the surface stress.

The main motivation of this work is to extend the study to a full micromagnetic framework in order to take into account spatial inhomogeneities and move a step forward the understanding of THz AFM dynamics driven by a damping-like torque originating from the spin Hall effect. We performed a systematic study of the threshold currents and the output frequency as a function of spin-polarization direction, exchange constant, Gilbert damping, AFM thickness and Dzyaloshinskii–Moriya interaction (DMI).

We found that the DMI is the most influent parameter in controlling the type of excited mode. At low DMI, the threshold current is a sub-critical Hopf bifurcation and the dynamics is related to a large amplitude uniform precession of the magnetization of the two sublattices in the same direction with a phase angle that depends on the applied current. As the DMI increases the ground state becomes non-uniform and the excited dynamics changes qualitatively and it is related to a continuous domain wall nucleation, propagation and annihilation. In addition, the threshold current is a super-critical Hopf bifurcation. Our results reveal the need of a full micromagnetic model for a proper design of AFM based oscillators.

This paper is organized as follows. Section II is devoted to the micromagnetic model developed for the analysis. Results are given in Section III in detailed paragraphs, then, the conclusions are summarized in Section IV.

II. MODEL

The device under examination consists of an AFM antiferromagnetic layer coupled to a 4 terminals Heavy metal layer [32] (see Fig. 1(a-b), where the Cartesian coordinate systems are also shown). The AFM has a square cross section with dimensions 40x40 nm$^2$, whereas the thickness $d$ varies from 1 to 5 nm. The continuous micromagnetic formalism used here extends the one of ferromagnets considering the macroscopic properties of an antiferromagnet as computed from averaging the spin vectors [39], in detail starting from the atomistic model the magnetization at each point is modeled by means of two vectors $\mathbf{m}_1$ and $\mathbf{m}_2$ (Fig. 1(c)) that are the average magnetic effect of the spins that point parallel or antiparallel a specific direction. AFM dynamics of $\mathbf{m}_1$ and $\mathbf{m}_2$ is obtained by solving two coupled Landau-Lifshitz-Gilbert equations, where the SHE-driven spin-transfer torque is taken into account by means of an additional Slonczewski-like torque term [36]:

$$\begin{cases} \dfrac{d\mathbf{m}_1}{d\tau} = -\left(\mathbf{m}_1 \times \mathbf{h}_{\text{eff-1}}\right) + \alpha \mathbf{m}_1 \times \dfrac{d\mathbf{m}_1}{d\tau} + d_J\left(\mathbf{m}_1 \times \mathbf{m}_1 \times \mathbf{p}\right) \\ \dfrac{d\mathbf{m}_2}{d\tau} = -\left(\mathbf{m}_2 \times \mathbf{h}_{\text{eff-2}}\right) + \alpha \mathbf{m}_2 \times \dfrac{d\mathbf{m}_2}{d\tau} + d_J\left(\mathbf{m}_2 \times \mathbf{m}_2 \times \mathbf{p}\right) \end{cases} \quad (1)$$

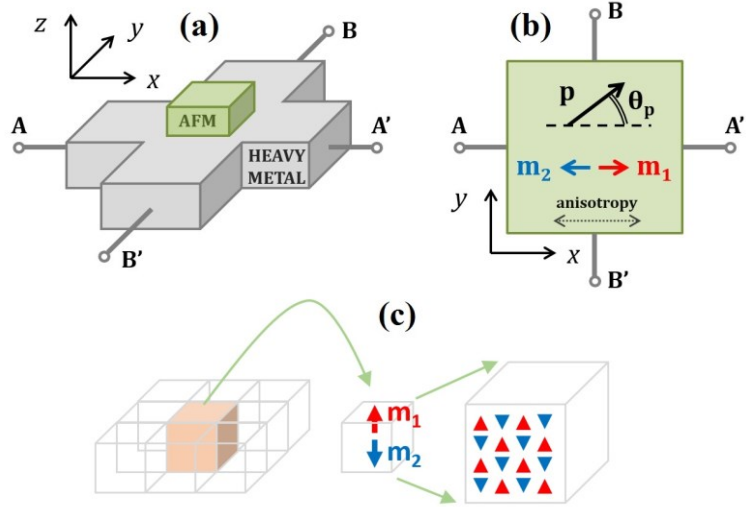

FIG. 1: Schemes of the device under investigation with the indication of the Cartesian reference systems. (a) A schematic of the bi-layered ASHO. The four terminals can be used for the application of charge currents, and for the measurement of the spin-Hall resistance. (b) Top view of the antiferromagnet, $\mathbf{m}_1$ and $\mathbf{m}_2$ represent the initial configuration of the magnetizations of the two sub-lattices while $\mathbf{p}$ is the spin polarization. (c) Sketch of the idea at the basis of the continuous modeling of antiferromagnetic sub-lattices, for a given computational cell we consider that the average magnetization is given by the two vectors $\mathbf{m}_1$ and $\mathbf{m}_2$.

In the left-hand side of Eqs. (1), $\mathbf{m}_1$ and $\mathbf{m}_2$ are therefore the magnetizations of the two sub-lattices, normalized with respect to the saturation magnetization $M_S$, and $\tau$ is the dimensionless time $\tau = \gamma_0 M_S t$, where $\gamma_0$ is the gyromagnetic ratio. In the right-hand side, $\mathbf{h}_{\text{eff-1}}$ and $\mathbf{h}_{\text{eff-2}}$ are the normalized effective fields acting on the two sub-lattices, and $\alpha$ is the Gilbert damping. The third term represents the SHE-driven torque, where $d_J = \dfrac{g \mu_B \theta_{SH} J}{2 \gamma_0 e M_S^2 d}$, $g$ is the Landè factor, $\mu_B$ is the Bohr magneton, $e$ is the electron charge, $\theta_{SH}$ is the spin-Hall angle, which represents the amount of charge current $J$ converted into spin current $J_S$, $\theta_{SH} = J_S/J$. The spin Hall effect is a Néel torque that is assumed to have the same form for each magnetic sublattice. The vector $\mathbf{p}$ is the direction of the spin-Hall polarization, given by $\mathbf{p} = \hat{z} \times \mathbf{j}$, where $\hat{z}$ and $\mathbf{j}$ are the directions of the spin and electric currents. By a proper combination of the current at the source terminals, it is possible to manage the direction of the spin-Hall polarization. In our case, $\mathbf{p}$ can be fixed in the x-y plane with an angle $\theta_\mathbf{p}$ between 0° and 90°: if the electric current is applied only at the terminals B-B' (A-A'), then $\theta_\mathbf{p} = 0°$ ($\theta_\mathbf{p} = 90°$) resulting in a polarization collinear (normal) to the easy axis, see Fig. 1(b).

The effective fields include the standard contributions from exchange, anisotropy, and demagnetizing:

$$\begin{cases} \mathbf{h}_{\text{eff-1}} = \mathbf{h}_{\text{exch-1}} + \mathbf{h}_{\text{ani-1}} + \mathbf{h}_{\text{demag-1}} \\ \mathbf{h}_{\text{eff-2}} = \mathbf{h}_{\text{exch-2}} + \mathbf{h}_{\text{ani-2}} + \mathbf{h}_{\text{demag-2}} \end{cases} \quad (2)$$

The exchange fields take into account both ferromagnetic coupling between neighbors in each sub-lattice (this is the same as in the standard model for the ferromagnets) and the antiferromagnetic coupling between the two sub-lattices. The latter is considered of atomistic origin because the two magnetization vectors are at the same point and it is modeled considering only the homogeneous part,

$$\begin{cases} \mathbf{h}_{\text{exch-1}} = \alpha_{\text{exch-FM}} \nabla^2 \mathbf{m}_1 + \lambda_{\text{AFM}} \mathbf{m}_2 \\ \mathbf{h}_{\text{exch-2}} = \alpha_{\text{exch-FM}} \nabla^2 \mathbf{m}_2 + \lambda_{\text{AFM}} \mathbf{m}_1 \end{cases} \quad (3)$$

where $\alpha_{\text{exch-FM}} = 2A_{FM}/\mu_0 M_S^2$ and $\lambda_{\text{AFM}} = 4A_{AFM}/\mu_0 a^2 M_S^2$ ponder the two main contributions, $A_{FM}$ and $A_{AFM}$ are the ferromagnetic and antiferromagnetic exchange constant, respectively, $a$ is the lattice constant, and $\mu_0$ is the vacuum permeability.

We consider anisotropy fields originating from uniaxial material:

$$\begin{cases} \mathbf{h}_{\text{ani-1}} = \alpha_{\text{ani}} \mathbf{m}_1 \cdot \mathbf{u}_k \\ \mathbf{h}_{\text{ani-2}} = \alpha_{\text{ani}} \mathbf{m}_2 \cdot \mathbf{u}_k \end{cases} \quad (4)$$

where $\alpha_{\text{ani}} = 2K_U/\mu_0 M_S^2$, $K_U$ is the uniaxial anisotropy constant, and $\mathbf{u}_K$ is the direction of the easy axis that is the *x* axis in our study.

The demagnetizing field is calculated by solving the magnetostatic problem [40] for the total magnetization $(\mathbf{m}_1 + \mathbf{m}_2)/2$. As we interested in the dynamics of ultra-thin AFM films, we assume here substantially low value of the homogeneous exchange $A_{AFM}/a^2 = 1.25$ MJ/m$^3$, being $a = 0.5$ nm. The discretization cell used for the simulations is 2nm x2nm x $d$. When not specified, we have used the following parameters for the ASHO: $d = 5$ nm, $M_S = 350 \times 10^3$ A/m, $\alpha = 0.05$, $K_U = 10^5$ J/m$^3$, $\theta_{SH} = 0.2$ and $A_{FM} = 0.5 \times 10^{-11}$ J/m.

## III. RESULTS

*A) Role of spin- polarization direction.*

We consider three experimental realizable spin-polarization directions $\mathbf{p}_1$, $\mathbf{p}_2$, and $\mathbf{p}_3$:

- $\mathbf{p}_1$ is obtained if the current is applied at the terminals B-B' along the -y direction $\theta_\mathbf{p} = 0°$, the spin polarization is collinear with the equilibrium magnetization of the two sub-lattices [30];
- $\mathbf{p}_2$ is obtained if the same current is applied simultaneously at both A-A' and B-B', $\theta_\mathbf{p} = 45°$ in the region where the AFM is positioned;
- $\mathbf{p}_3$ is obtained if the electric current is applied at the terminals A-A' along the x direction hence $\theta_\mathbf{p} = 90°$, the spin polarization is perpendicular to the equilibrium magnetization of the two sub-lattices [32].

Fig. 2(a) and (b) show the threshold currents and the oscillation frequencies as a function of current density for the three different spin polarizations without DMI. In all the cases, the self-oscillation is a sub-critical Hopf bifurcation characterized by hysteresis with $J_{ON}$ and $J_{OFF}$ switching on and switching off current densities respectively. This hysteretic excitation has been already predicted by an analytical theory for the $\mathbf{p}_3$ configuration [32] and can be understood qualitatively by considering that at $J_{ON}$ the magnetization precession of the magnetization of the two sub-lattices has a finite large amplitude. Differently from sub-critical Hopf bifurcation in ferromagnet-based spin-transfer torque oscillators [17, 41, 42, 43], here also at $J_{OFF}$ the oscillation amplitude is finite and even larger than the one at $J_{ON}$ (see Fig. 2(c) and (d) where the amplitude of the y-component of the magnetization for $\theta_\mathbf{p} = 0°$ and 90° as a function of current density is displayed). This result is relevant from a technological point of view because the AFM oscillator can also work at a current density below $J_{ON}$ as already pointed out in [32]. The width of the hysteretic region depends on the polarization direction as for $\theta_\mathbf{p} = 0°$ it is very narrow ($0.4 \times 10^8$ A/cm$^2$), whereas it increases with the increase of $\theta_\mathbf{p}$ ($3.2 \times 10^8$ A/cm$^2$ for $\theta_\mathbf{p} = 90°$). This result can be directly linked to the fact that the precession axis is parallel to the spin polarization, then at $\theta_\mathbf{p} = 0°$ it coincides with the equilibrium axis while at $\theta_\mathbf{p} = 90°$ the precession axis is perpendicular to it (see top right inset of Fig. 3).

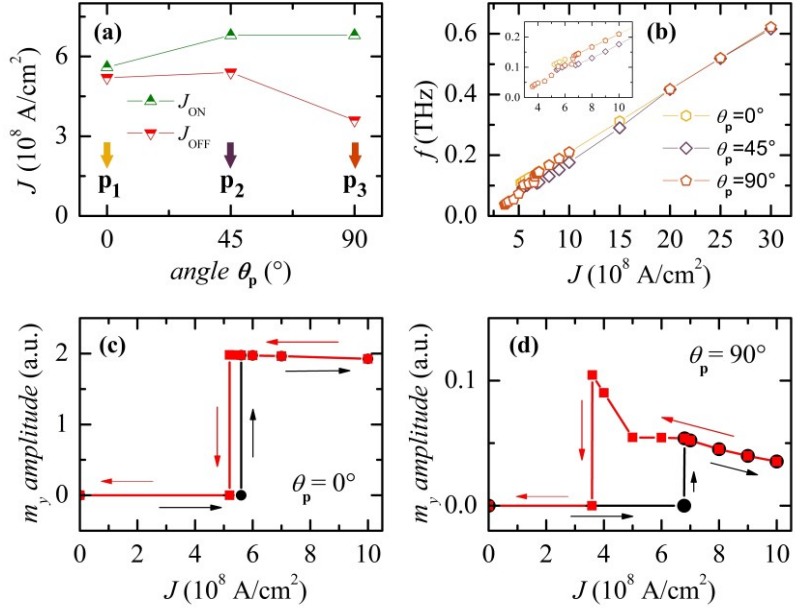

FIG. 2: (a) Threshold current densities ($J_{ON}$ and $J_{OFF}$) for the excitation of the antiferromagnetic dynamics for three different directions of spin-polarization. (b) Oscillation frequency as a function of the applied current with a zoom near the threshold current. (c)-(d) Amplitude of the *y*-component of the magnetization as a function of the current density for $\theta_p = 0°$ and $\theta_p = 90°$.

The AFM magnetization dynamics is characterized by the rotation of the magnetization of both sublattices $\mathbf{m}_1$ and $\mathbf{m}_2$ in the same direction with a phase angle $\psi$ with respect to the oscillation axis (top left inset of Fig. 3). The rotation frequency (Fig. 2(b)) exhibits blue shift tunability (21GHz/($10^8$A/cm$^2$)) and it turns out that oscillation frequencies (above $J_{ON}$) are basically independent of the spin-polarization direction. The origin of this result is given by the fact that for a fixed current density the trajectory is characterized by the same $\psi$ around the oscillation axis fixed by the spin-polarization direction. This fact is preserved also at very large current, see for example the main panel of Fig. 3 for the trajectories at $J = 30 \times 10^8$ A/cm$^2$. As expected from analytical computations the frequency is proportional to the current density (see Eq. 7 of Ref. [32]). For the simulation parameters of this study, a maximum frequency of 0.6 THz at $J = 30 \times 10^8$ A/cm$^2$ is observed.

At $\theta_p = 90°$, we have performed a comparison with the analytical model developed in Ref. [32] finding an agreement, described below in the paper (see also Supplementary Note 1 [44]).

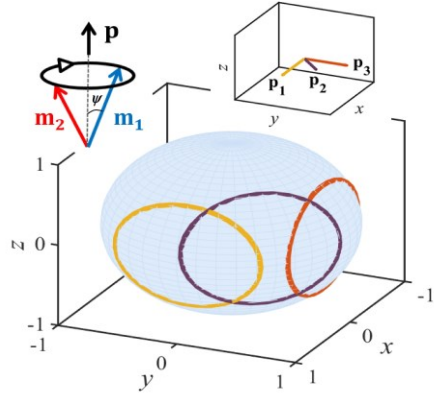

FIG. 3: Trajectories of the magnetizations of the two sub-lattices in the three different cases of spin-Hall polarization, around its direction, for $J = 30 \times 10^8$ A/cm$^2$. Left inset: sketch of the precession of the two magnetizations around the spin polarization. Right inset: directions of the spin-polarization in the three cases.

*B) Output signal*

The first challenge to face is the conversion of the AFM dynamics in a measureable THz signal. Some proposed strategies are based on the inverse spin-Hall effect [32] or dipolar radiation [45]. Those two approaches need tilting of the magnetization of the two sublattices for originating a net rotating magnetic vector or a time varying phase angle between the two sublattices, however for realistic parameters the output power should be very small. On the other hand, our 4-terminal scheme can be used biasing the device with a proper current in order to have $\mathbf{p}_1$, $\mathbf{p}_2$ and $\mathbf{p}_3$, and reading the magnetoresistance at one of the couples of terminals AA' or BB' [46, 47]. For example when the bias current is applied through the AA' terminals and hence the spin-polarization is $\mathbf{p}_3$, the THz signal should be read out via the BB' terminals and it is mainly originated by the oscillation of the *y*-component of the magnetization of the two sublattices; such an oscillation has a frequency that is two times the precession frequency (see Supplementary Note 2 [44]). Alternatively, the THz signal can be read via the same AA' terminal via the magnetoresistance that originates from the oscillation of the *x*-component of the magnetization of the two sublattices [48].

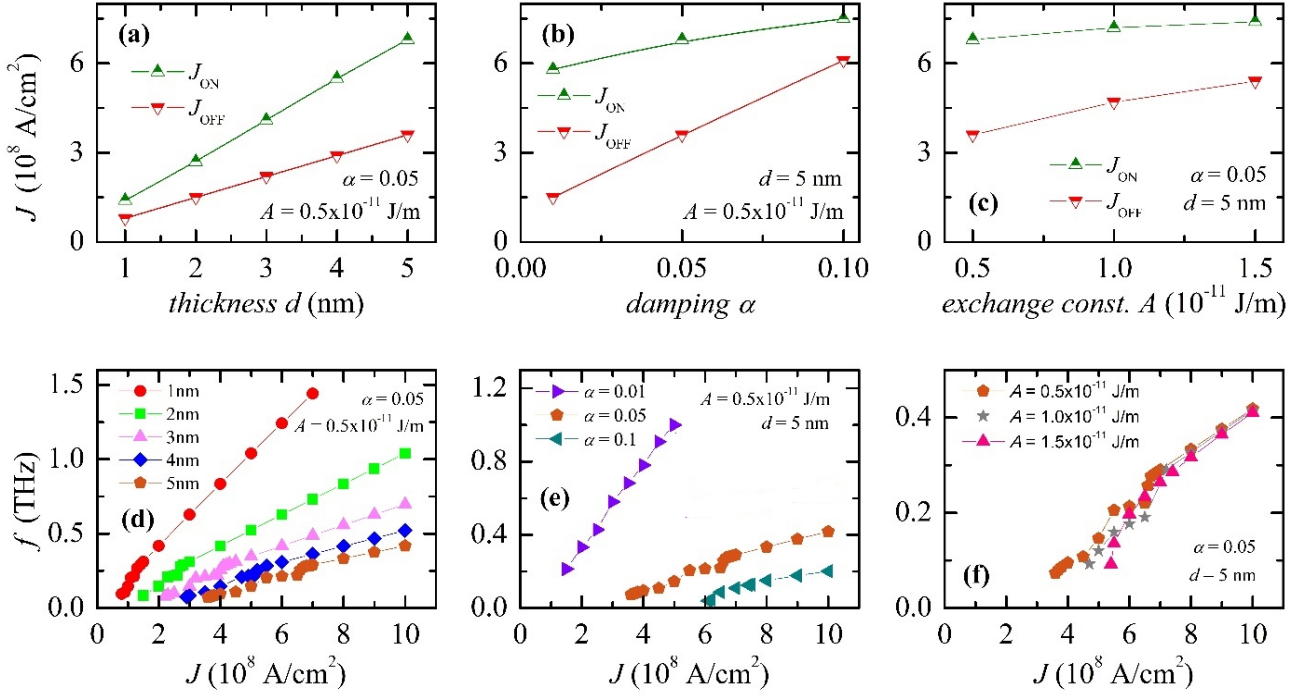

FIG. 4: Summary of micromagnetic simulations for a current applied along the *x*-axis, so that the spin-Hall polarization is along the *y*-axis ($\theta_p = 90°$). (a)-(c) Switching on and off current densities to as a function of AFM thickness *d* (a), damping *α* (b), and exchange constant *A* (c). (d)-(f) Oscillation frequency of the *y*-component of the magnetization of the AFM as a function of the current density, for different values of the thickness *d* (d), the damping *α* (e), and the exchange constant *A* (f).

C) *Systematic study for* $\mathbf{p}_3$ *spin-polarization*.

Figs. 4(a)-(c) show the switching on $J_{ON}$ and switching off $J_{OFF}$ current density as a function of *d*, *α*, and *A* while maintaining the other two parameters constant. The threshold currents clearly increase with the increase of both the AFM thickness and the damping (Figs. 4(a)-(b)). On the other hand, our simulations confirm that the exchange contribution plays an important role mainly in the switching off current density, which slightly increases with the value of *A*, whereas the switching on current density is almost constant (Fig. 4(c)). The hysteresis width increases with the thickness, decreases with the damping, and slightly decreases with the value of *A*. Such results agree with the theoretical predictions (see Eqs. (4) and (5) of Ref. [32]).

Within the same parametric study, Figs. 4(d)-(f) show the oscillation frequency (as computed from the *y*-component of the magnetization) of the excited dynamics as a function of the applied current *J*, for different values of thickness, damping and exchange constant. The frequency tuneability is blue-shift on current with frequency values ranging from hundreds of GHz up to several THz. In particular, the frequencies increase with either the decrease of thickness (Fig. 4(d)) and damping (Fig. 4(e)). In conclusion, full numerical micromagnetic simulations are in qualitative agreement with the theoretical predictions that hence can be used as a first tool to identify the parameter region where to optimize the THz AFM-based oscillators (see Fig. 4(f) and Eqs. (6) and (7) of Ref. [32]).

The oscillation frequencies in Fig 4(e), computed for $\alpha$=0.01, shows a jump to zero at $J = 5 \times 10^8$ A/cm$^2$ where the dynamics of the y-component of the magnetization is off ($m_y$ is constant) and the trajectory is in the x-z plane. This is a direct consequence of the reduced exchange stiffness or, in other words, the low thickness of the AFM film. As can be observed the damping is a critical parameter either for the oscillation frequency and for the range of current tunability. This brings us to the conclusion that the THz dynamics in ultra low damping AFM materials will be observable in a narrow range of current density, at least if we readout the signal via the spin-Hall resistance.

*D) ASHO linewidth.*

Together with frequency tunability and threshold current, the linewidth is another fundamental property of an oscillator. In order to calculate the linewidth for the AFM oscillator, we have performed micromagnetic simulations at room temperature (*T*=300 K) by considering the thermal field, as a stochastic contribution added to the deterministic effective field:

$$\begin{bmatrix} \mathbf{h}_{TH-1} \\ \mathbf{h}_{TH-2} \end{bmatrix} = \frac{\xi}{M_S} \sqrt{\frac{2\alpha k_B T}{\mu_0 \gamma_0 \Delta V M_S \Delta t}} \quad (6)$$

with $k_B$ is the Boltzmann constant, $\Delta V$ and $\Delta t$ are the discretization volume and integration time step, respectively, while *T* is the temperature. $\xi$ is a 6-dimentional white Gaussian noise with zero mean and unit variance, uncorrelated for each computational cell [49].

We have computed the linewidth for different values of current density, *T*=300 K, $\theta_p = 90°$, and the default values for *d*, α, and *A*. Our results point out that it is smaller than 10 MHz (our simulations are 100ns long), corresponding to a quality factor of $Q = f/\Delta f = 41000$ at least.

*E) Comparison with analytical model.*

As already cited, our main numerical results agree with recently published theoretical predictions [32]. For this reason, we focused on a direct comparison between micromagnetic simulations and those analytical models, finding a good agreement for both the threshold currents and the output frequencies. Fig. 5 summarizes this comparison. In the first graph, numerical threshold currents, as a function of the AFM layer thickness, are compared with the analytical formulas (Eqs. (4) and (5) in Ref. [32]):

$$\begin{cases} J_{ON} = \dfrac{\omega_{ani}}{2\sigma} \\ J_{OFF} = \dfrac{2\alpha}{\pi\sigma}\sqrt{\omega_{exch}\omega_{ani}} \end{cases} \quad (7)$$

where $\omega_{ani} = \gamma_0(2K_U/M_S)$, $\sigma = (g\mu_B\theta_{SH}/2eM_Sd)$, $\omega_{exch} = \gamma_0(4A_{AFM}/a^2M_S)$.

Fig. 5(b) shows the comparison concerning the output frequency of the oscillator for the default set of parameters. The analytical formula corresponds to Eq. (7) of Ref. [32]:

$$\omega = \dfrac{\sigma J}{\alpha} \quad (8)$$

where, however, we are referring to the double frequency of the $y$-component of the magnetization.

We also performed numerical and analytical calculations in the case of higher exchange, considering $A_{AFM}/a^2 = 20$ MJ/m$^3$. Again, the comparison is convincing (see Fig. 5(c)), and we can state that, from the qualitative point of view, there is no significant change in the dynamics and in the inertial nature of their excitation.

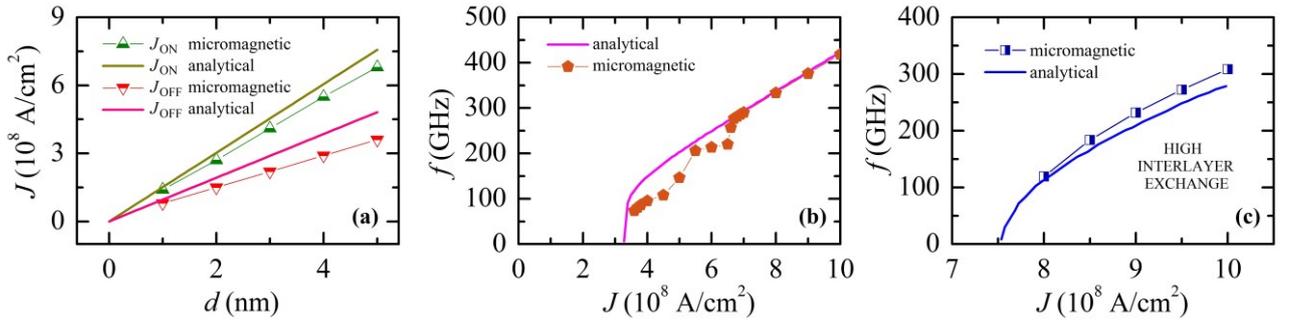

FIG. 5: Comparison between micromagnetic simulations and analytical models in the case $\theta_p = 90°$: (a) threshold currents, (b) oscillation frequency of the $y$-component of the magnetization, (c) oscillation frequency of the $y$-component of the magnetization in the case of high interlayer exchange ($A_{AFM}/a^2 = 20$ MJ/m$^3$).

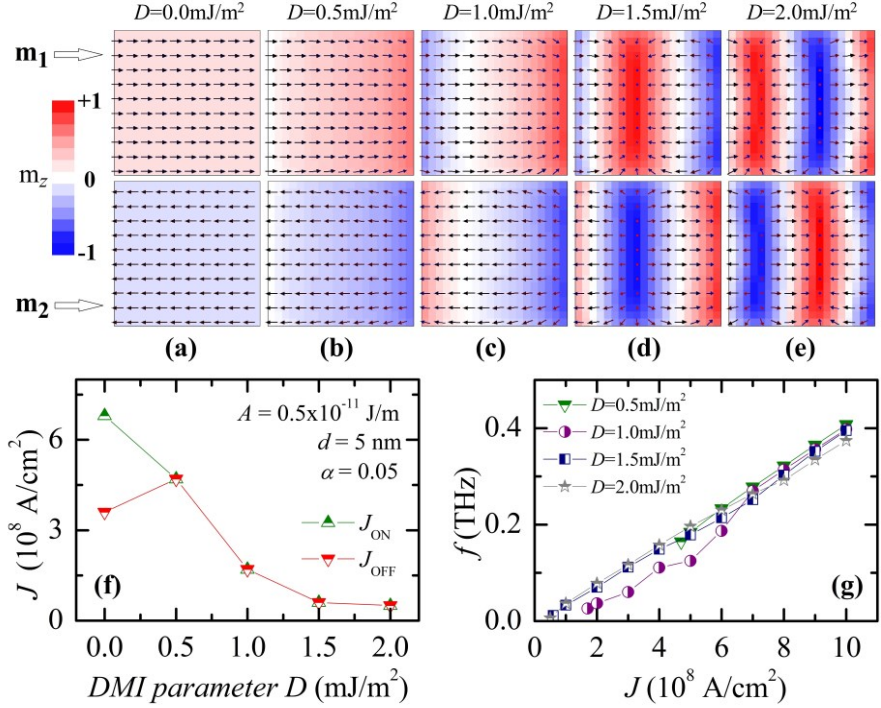

FIG. 6: (a)-(e) Equilibrium configurations of the magnetization in the two sub-lattices as a function of the interfacial DMI parameter $D$. (f) Switching on and off current densities as a function of $D$. (g) Oscillation frequency of the spin-Hall magnetoresistance as a function of current density for different values of $D$.

*F) Effect of the DMI.* The need of the full micromagnetic framework to analyze the magnetization dynamics in an AFM driven by SHE is clear in presence of the interfacial DMI. This effect is taken into account by considering an additional contribution to the effective field having the following expression:

$$\begin{cases} \mathbf{h}_{\text{DMI-1}} = -\dfrac{2D}{\mu_0 M_S}\left[(\nabla \cdot \mathbf{m}_1)\hat{z} - \nabla m_{z\text{-}1}\right] \\ \mathbf{h}_{\text{DMI-2}} = -\dfrac{2D}{\mu_0 M_S}\left[(\nabla \cdot \mathbf{m}_2)\hat{z} - \nabla m_{z\text{-}2}\right] \end{cases} \qquad (5)$$

with $D$ being the parameter accounting for the intensity of DMI. The boundary conditions now hold $\dfrac{d\mathbf{m}_i}{dn} = \dfrac{1}{\chi}(\hat{z}\times\mathbf{n})\times\mathbf{m}_i$ ($i$=1,2), where $\mathbf{n}$ is the unit vector perpendicular to the edge and $\chi = \dfrac{2A_{FM}}{D}$ is a characteristic length in presence of DMI. The first effect of the interfacial DMI is on the ground state. In particular, Fig. 6 shows the evolution of the equilibrium configuration of the magnetization for different $D$. Starting from the uniform state (Fig. 6(a)), Nèel-type DWs are stabilized starting from $D=1.5$mJ/m$^2$ (see Figs. 6(c)-(e)) [50, 51, 52]. The second effect is the change of the bifurcation at $D=1.0$mJ/m$^2$ from sub-critical to super-critical and hence with the disappearing of the hysteretic excitation ($J_{ON}=J_{OFF}$) as displayed in Fig. 6(f). The third effect is the qualitative change of the magnetization dynamics that now it is characterized by a continuous nucleation, shifting and annihilation of DWs along a direction that depends on the applied current (see Supplemental Material [44], Movies 1 and 2, to compare the dynamics at $D=0.0$mJ/m$^2$ and $D=2.0$mJ/m$^2$) [53]. Fig. 6(g) summarizes the output frequency as a function of current density for different DMI parameter and it turns out that DMI does not play a very important role in this case. This result is due to the fact that the main role of the DMI is the stabilization of the domain wall chirality.

We performed micromagnetic simulations considering a high interlayer exchange, also in the case of interfacial DMI. The magnetic configuration of the AFM sublattices is still characterized by non-uniform DWs, which translate along the current as in the case of low exchange (see Fig. 7(a) with the equilibrium configuration of sublattices obtained for $D=2.0$mJ/m$^2$). The high exchange, moreover, does not influence significantly the frequency of dynamics, as shown in Fig. 7(b).

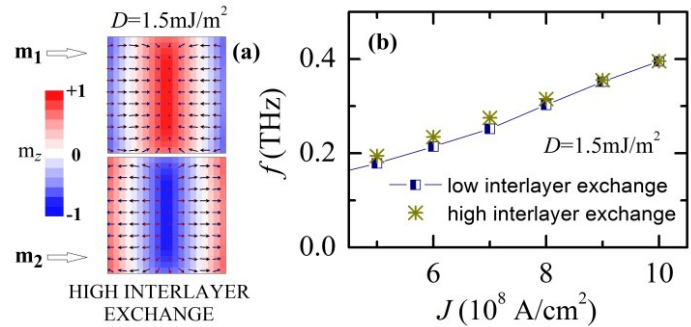

FIG. 7: (a) Equilibrium configuration of the magnetization in the two sub-lattices in the case of high interlayer exchange ($A_{AFM}/a^2 = 20$ MJ/m$^3$) for $D=1.5$mJ/m$^2$. (b) Comparison of the output frequency with low and high interlayer exchange for $D=1.5$mJ/m$^2$.

III. CONCLUSIONS

AFM materials are promising for the realization of a compact sub micrometer scale THz oscillator tunable with a current in a wide range of frequency ranging from few hundreds of GHz up to 1-2 THz. Actually, this idea is still not demonstrated experimentally, this paper contributes to furnish a more detailed numerical understanding of the THz dynamics driven by spin-Hall effect. We find that the macrospin based theoretical model can be used for a qualitative study at very low DMI while a full micromagnetic approach is necessary in presence of DMI, that is an energy contribution that arises in most of the experimental promising solutions for AFM based oscillators. In conclusion, we point out that the interlayer exchange interaction within the same computational cell has no influence on the domain wall nucleation being a local term only.

*Acknowledgments.* The authors thank Takahiro Moriyama, Oksana Fesenko and Pedram Khalili Amiri for the fruitful discussions.

REFERENCES

[1] P. H. Siegel, IEEE Trans. Microwave Theory Tech. **50**, 910 (2002).
[2] B. Ferguson and X.-C. Zhang, Nature Mater. **1**, 26 (2002).
[3] C. Kulesa, IEEE Trans. THz Science and Techn. **1**, 232 (2011).
[4] B. Fisher, M. Hoffmann, H. Helm, G. Modjesch, and P.U. Jepsen, Semicond. Science and Techn. **20**, S246 (2005).
[5] M.C. Beard, G.M. Turner, and C.A. Schmuttenmaer, Phys. Med. Biol. **47**, 3841 (2002).
[6] J.F. Federici, B. Schulkin, F. Huang, D. Gary, R. Barat, F. Oliveira, and D. Zimdars, Semicond. Science and Techn. **20**, S266 (2005).
[7] T.J. Yen, W.J. Padilla, N. Fang, D.C. Vier, D.R. Smith, J.B. Pendry, D.N. Basov, and X. Zhang, Science **303**, 1494 (2004).
[8] C.D. Stoik, M.J. Bohn, and J.L. Blackshire, Optics Expr. **16**, 17039 (2008).
[9] M. Tonouchi, Nature Photon. **1**, 97 (2007).
[10] W. Withayachumnankul, G.M. Png, and X. Yin, Proc. IEEE **95**, 1528 (2007).
[11] B. S. Williams, Nature Photon. **1**, 517 (2007).
[12] E. Seok, D. Shim, C. Mao, R. Han, S. Sankaran, C. Cao, W. Knap, and K.K. O, IEEE Journ. Solid-State Circ. **45**, 1554 (2010).
[13] M. Tsoi, A. G. M. Jansen, J. Bass, W.-C. Chiang, M. Seck, V. Tsoi, and P. Wyder, Phys. Rev. Lett. **80**, 4281 (1998).
[14] W. H. Rippard, M. R. Pufall, S. Kaka, S. E. Russek, and T. J. Silva, Phys. Rev. Lett. **92**, 027201 (2004).
[15] V. Puliafito, B. Azzerboni, G. Consolo, G. Finocchio, L. Torres, and L. Lopez-Diaz, IEEE Trans. Magn. **44**, 2512 (2008).
[16] Z. Zeng, G. Finocchio, and H. Jiang, Nanoscale **5**, 2219 (2013).
[17] A. Giordano, R. Verba, R. Zivieri, A. Laudani, V. Puliafito, G. Gubbiotti, R. Tomasello, G. Siracusano, B. Azzerboni, M. Carpentieri, A. Slavin, and G. Finocchio, Sci Rep. **6**, 36020 (2016).
[18] V. Puliafito, A. Giordano, A. Laudani, F. Garescì, M. Carpentieri, B. Azzerboni, and G. Finocchio, Appl. Phys. Lett. **109**, 202402 (2016).
[19] L. Duo, M. Finazzi, and F. Ciccacci, Magnetic properties of antiferromagnetic oxide materials: surfaces, interfaces, and thin films, Weinheim: Wiley-VCH Verlag GmbH & Co (2010).
[20] V. Baltz, G. Gaudin, P. Somani, and B. Dieny, Appl. Phys. Lett. **96**, 262505 (2010).
[21] S.-H. Yang, K.-S. Ryu, and S. Parkin, Nat. Nanotechnol. **10**, 221 (2015).
[22] R. Tomasello, V. Puliafito, E. Martinez, A. Manchon, M. Ricci, M. Carpentieri, and G. Finocchio, J. Phys. D: Appl. Phys. **50**, 325302 (2017).
[23] O. Gomonay, K. Yamamoto, and J. Sinova, J. Phys. D: Appl. Phys. **51**, 264004 (2018).
[24] P.E. Roy, R.M. Otxoa, and J. Wunderlich, Phys. Rev. B **94**, 014439 (2016).
[25] K. Olejnik, V. Schuler, X. Marti, V. Novák, Z. Kaspar, P. Wadley, R.P. Campion, K.W. Edmonds, B.L. Gallagher, J. Garces, M. Baumgartner, P. Gambardella, and T. Jungwirth, Nature Comm. **8**, 15434 (2017).
[26] A. S. Núñez, R. A. Duine, P. Haney, and A. H. MacDonald, Phys. Rev. B **73**, 214426 (2006).
[27] T. Moriyama, N. Matsuzaki, K.-J. Kim, I. Suzuki, T. Taniyama, and T. Ono, Appl. Phys. Lett. **107**, 122403 (2015).

[28] L. Liu, T. Moriyama, D. C. Ralph, and R. A. Buhrman, Phys. Rev. Lett. **106**, 036601 (2011).

[29] E. V. Gomonay and V. M. Loktev, Low Temp. Phys. **40**, 17 (2014).

[30] R. Cheng, D. Xiao, and A. Brataas, Phys. Rev. Lett. **116**, 207603 (2016).

[31] S. Baierl, J.H. Mentink, M. Hohenleutner, L. Braun, T.-M. Do, C. Lange, A. Sell, M. Fiebig, G. Woltersdorf, T. Kampfrath, and R. Huber, Phys. Rev. Lett. **117**, 197201 (2016).

[32] R. Khymyn, I. Lisenkov, V. Tiberkevich, B. A. Ivanov, and A. Slavin, Sci. Rep. **7**, 43705 (2017).

[33] V. Baltz, A. Manchon, M. Tsoi, T. Moriyama, T. Ono, and Y. Tserkovnyak, Rev. Mod. Phys. **90**, 015005 (2018).

[34] O. Gomonay, V. Baltz, A. Brataas, and Y. Tserkovnyak, Nature Phys. **14**, 213 (2018).

[35] M.B. Jungfleish, W. Zhang, and A. Hoffmann, Phys. Lett. A **382**, 865 (2018).

[36] A. Manchon, I.M. Miron, T. Jungwirth, J. Sinova, J. Zelezny, A. Thiaville, K. Garello, and P. Gambardella, arXiv:1801.09636 (2018).

[37] N. Ntallis and K.G. Efthimiadis, Comp. Mat. Science **97**, 42 (2015).

[38] D. Alders, L. H. Tjeng, F. C. Voogt, T. Hibma, G. A. Sawatzky, C. T. Chen, J. Vogel, M. Sacchi, and S. Iacobucci, Phys. Rev. B **57**, 11623 (1998).

[39] D. Suess, T. Schrefl, W. Scholz, J.-V. Kim, R. L. Stamps, and J. Fidler, IEEE Trans. Magn. **38**, 2397 (2002).

[40] L. Lopez-Diaz, D. Aurelio, L. Torres, E. Martinez, M. A. Hernandez-Lopez, J. Gomez, O. Alejos, M. Carpentieri, G. Finocchio, and G. Consolo, J. Phys. D: Appl. Phys. **45**, 323001 (2012).

[41] M. D'Aquino, C. Serpico, R. Bonin, G. Bertotti, and I.D. Mayergoyz, Phys. Rev. B **84**, 214415 (2011).

[42] G. Finocchio, V. Puliafito, S. Komineas, L. Torres, O. Ozatay, T. Hauet, and B. Azzerboni, J. Appl. Phys. **114**, 163908 (2013).

[43] V. Puliafito, Y. Pogoryelov, B. Azzerboni, J. Akerman, and G. Finocchio, IEEE Trans. Nanotechnol. **13**, 532 (2014).

[44] See Supplemental Material at [ ] for more information about the inertial nature of the antiferromagnetic oscillator, the double frequency of the component of the magnetization along the spin-polarization, and for the movies described in the text.

[45] O. R. Sulymenko, O. V. Prokopenko, V. S. Tiberkevich, A. N. Slavin, B. A. Ivanov, and R. S. Khymyn, Phys. Rev. Appl. **8**, 064007 (2017).

[46] J. Sinova, S. O. Valenzuela, J. Wunderlich, C. H. Back, and T. Jungwirth, Rev. Mod. Phys. **87**, 1213 (2015).

[47] T. Moriyama, K. Oda, and T. Ono, arXiv:1708.07682v1 (2017).

[48] S.Yu. Bodnar, L. Smeikal, I. Turek, T. Jungwirth, O. Gomonay, J. Sinova, A.A. Sapozhnik, H.-J. Elmers, M. Klaui, and M. Jourdan, Nature Communications **9**, 348 (2018).

[49] G. Finocchio, I.N. Krivorotov, X. Cheng, L. Torres, and B. Azzerboni, Phys. Rev. B **83**, 134402 (2011).

[50] S. Emori, U. Bauer, S.-M. Ahn, E. Martinez, and G.S.D. Beach, Nat. Mater. **12**, 611 (2013).

[51] G. Siracusano, R. Tomasello, A. Giordano, V. Puliafito, B. Azzerboni, O. Ozatay, M. Carpentieri, and G. Finocchio, Phys. Rev. Lett. **117**, 087204 (2016).

[52] V. Puliafito, A. Giordano, B. Azzerboni, and G. Finocchio, J. Phys. D: Appl. Phys. **49**, 145001 (2016).

[53] M. Cubukcu, J. Sampaio, K. Bouzehouane, D. Apalkov, A. V. Khvalkovskiy, V. Cros, and N. Reyren, Phys. Rev. B **93**, 020401 (2016).

# Micromagnetic modeling of Terahertz oscillations in an antiferromagnetic material driven by spin-Hall effect

## SUPPLEMENTAL MATERIAL

V. Puliafito[1], R. Khymyn[2], M. Carpentieri[3], B. Azzerboni[1], V. Tiberkevich[4], A. Slavin[4], G. Finocchio[5]

[1] Department of Engineering, University of Messina, 98166 Messina, Italy
[2] Department of Physics, Gothenburg University, 40530 Gothenburg, Sweden
[3] Department of Electrical and Information Engineering, Politecnico of Bari, 70125 Bari, Italy
[4] Department of Physics, Oakland University, 48309 Rochester, MI, United States
[5] Department of Mathematical and Computer Sciences, Physical Sciences and Earth Sciences, University of Messina, 98166 Messina, Italy

**Supplementary Note 1**

*The inertial nature of the antiferromagnetic oscillator.* In the case of spin-Hall polarization perpendicular to the easy axis, namely for $\theta_p = 90°$, we have obtained a good agreement with the results of the analytical model developed in Ref. [1], as shown in the main text. In particular, this agreement concerns the presence of two threshold currents for the oscillations, an "ignition" current $J_{ON}$ and an elimination current $J_{OFF}$, the latter lower than the former so highlighting an inertial nature of the excitation. Fig. S1 is a representation of this feature, and it aims to reproduce, qualitatively, Fig. 4 of Ref. [1]. We show two abrupt steps of DC current applied to the heavy metal of the antiferromagnetic spin-Hall oscillator in the default configuration of the parameters. One step starts from value corresponding to the "ignition" current, the second step starts from zero. Both of them, however, finish at value slightly higher than the elimination current. We also show the oscillation of the *y*-components of the magnetization of one sub-lattice, corresponding to those steps of current. Although both the steps finish at the same working value, in the first case we get a persistent oscillation, in the second there is no oscillation. From a qualitative point of view, this behaviour is similar to what shown in Ref. [1].

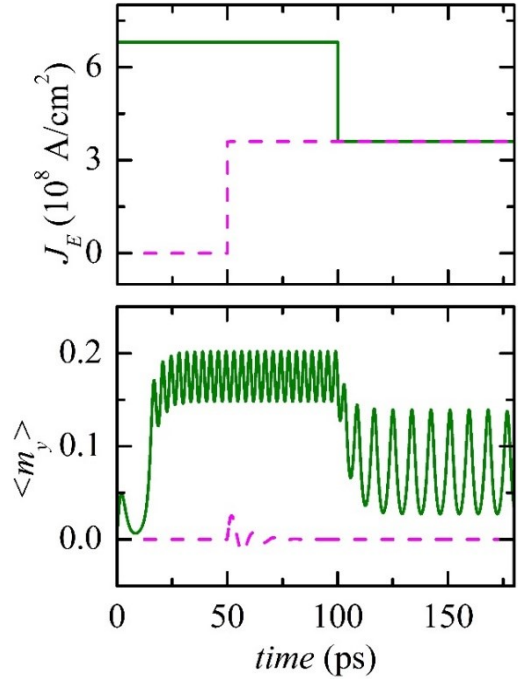

FIG. S1: (a) Two different temporal evolutions of the applied current in the heavy metal and (b) corresponding dynamics of the *y*-component of the magnetization in one of the two sub-lattices.

**Supplementary Note 2**

*The double frequency of the component of the magnetization along the spin-polarization.* If we consider, again, the case of $\theta_p = 90°$, the trajectory of the two sublattices turns out non-planar (Fig. S2a), as also obtained by the theoretical approach of Ref. [1]. This is basically due to the trade-off between the polarization and the easy axis which are perpendicular each other. This non-planar trajectory is almost negligible in the other two cases of polarization, where all the components oscillate at the same frequency (Fig. S2b). In the case of $\theta_p = 90°$, therefore, the *y*-component of the magnetization oscillates at a double frequency with respect to the frequency of precession (Fig. S2c). On the other hand, the amplitude of this oscillation of the *y*-component turns out quite small (see again Fig. S2a) and consequently also output power is low. For this reason, if a higher power is necessary we should consider to read the output signal via the magnetoresistance that originates from the oscillation of the *x*-component of the magnetization of the two sublattices [2, 3].

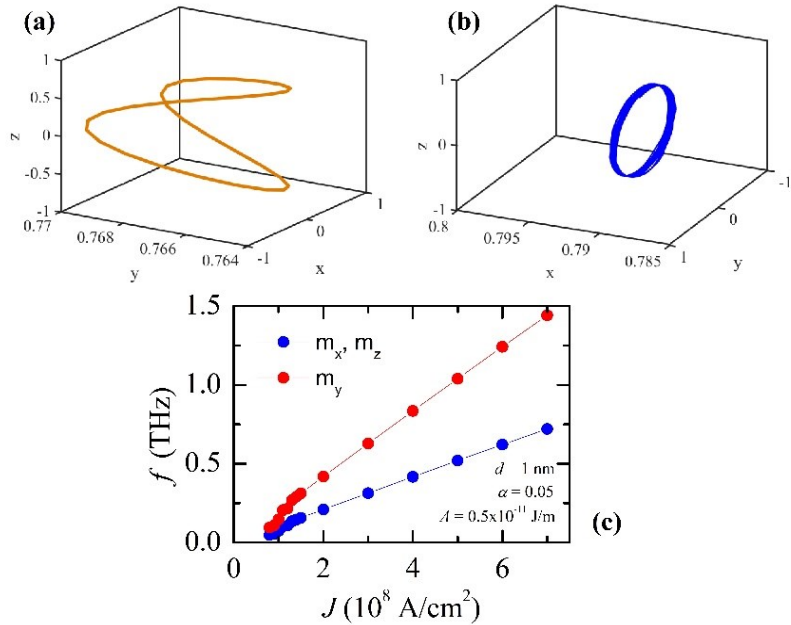

FIG. S2: Trajectory of the magnetization of the sub-lattices for (a) $\theta_p = 90°$ and (b) $\theta_p = 0°$. (c) Example of frequencies of the three components of the magnetization in a sub-lattice when $\theta_p = 90°$ (spin-Hall polarization along *y*) with the component of the magnetization along the spin-polarization having double frequency.

REFERENCE

[1] R. Khymyn, I. Lisenkov, V. Tiberkevich, B. A. Ivanov, and A. Slavin, Sci. Rep. **7**, 43705 (2017).
[2] T. Moriyama, K. Oda, and T. Ono, arXiv:1708.07682v1 (2017).
[3] S.Yu. Bodnar, L. Smeikal, I. Turek, T. Jungwirth, O. Gomonay, J. Sinova, A.A. Sapozhnik, H.-J. Elmers, M. Klaui, and M. Jourdan, Nature Communications **9**, 348 (2018).